\documentclass[twocolumn,showpacs,preprintnumbers,amsmath,amssymb]{revtex4}
\usepackage{graphicx}% Include figure files
\usepackage{dcolumn}% Align table columns on decimal point
\usepackage{bm}% bold math
\usepackage{color}% color
\usepackage{float}

\begin{document}

%\preprint{APS/123-QED}

\title{Investigation of neutron transfer in $^{7}$Li+$^{124}$Sn reaction}% Force line breaks with \\

\author{V. V. Parkar$^{1,2}$ \footnote{vparkar@barc.gov.in}}
\author{A. Parmar$^3$}
\author{Prasanna M.$^4$}
\author{V. Jha$^{1,2}$}
\author{S. Kailas$^5$}
\affiliation{$^1$Nuclear Physics Division, Bhabha Atomic Research Centre, Mumbai - 400085, India}
\affiliation{$^2$Homi Bhabha National Institute, Anushaktinagar, Mumbai - 400094, India}
\affiliation{$^3$Sardar Vallabhbhai National Institute of Technology, Surat - 395007, India}
\affiliation{$^4$Department of Physics, Rani Channamma University, Belagavi - 591156, India}
\affiliation{$^5$UM-DAE Centre for Excellence in Basic Sciences, Mumbai-400098, India}


\begin{abstract}
The relative importance of neutron transfer and breakup process in reaction around Coulomb barrier energies have been studied for the $^{7}$Li+$^{124}$Sn system. Coupled channel calculations have been performed to understand the one neutron stripping and pickup cross sections along with the breakup in the $^{7}$Li+$^{124}$Sn system. The systematics of one and two neutron striping and pickup cross sections with $^7$Li projectile on several targets show an approximate universal behaviour that have been explained by a simple model. Complete reaction mechanism have been studied by comparing the reaction cross sections with cumulative cross sections of total fusion and one neutron transfer.  
\end{abstract}

\maketitle

\section{\label{sec:Intro} Introduction}
Weakly bound nuclei are characterized by dominant cluster structures and a loose binding with respect to the breakup into these clusters. These features are linked to enhanced cross sections of breakup and transfer channels in reactions around Coulomb barrier energies using weakly bound projectiles (WBP). The investigations into the relative importance of  different processes on the reaction mechanism is a topic of intense current interest. In this context, several experimental studies have been performed over the  years utilizing projectiles of both stable and unstable weakly bound nuclei. Various processes such as, elastic scattering, complete and incomplete fusion, inclusive and exclusive  breakup, transfer have been studied in reactions around Coulomb barrier energies using WBP \cite{Cant15, Jha20}. In particular, the stable WBPs such as,  $^6$Li, $^7$Li and $^9$Be on several targets have been extensively used for such measurements. Many new features have been highlighted from these studies, which were not observed with strongly bound projectiles (SBP).

From the measurements of elastic scattering, the total reaction cross sections with WBP are found to be much larger than those with comparative SBP \cite{Kola16, Aguilera2011}. Further, a new type of anomaly in the optical potential description of elastic scattering, namely the `breakup threshold anomaly' is observed in the case of WBPs, which has been usually attributed to the repulsive polarization potential produced by the cluster breakup process of projectile \cite{Kumawat20}. In the studies of fusion with WBPs, the complete fusion (CF) cross sections are found to be suppressed at above barrier energies with respect to predictions of one-dimensional barrier penetration model (1DBPM) \cite{Cant15, Jha20}. An interesting observation in these measurements is that the amount of suppression is commensurate with the measured incomplete fusion (ICF) \cite{VVP18, VVP18b}. Total fusion (TF) which is sum of CF and ICF cross sections match with the 1DBPM predictions at above barrier energies \cite{Jha20}. A large inclusive $\alpha$-particle cross sections measured at energies around the Coulomb barrier is another fascinating feature of reactions with WBP \cite{Kumawat10, Santra12, Pandit17, Jha20}. In the breakup measurements, it was observed that the non-capture breakup (NCBU) is very small compared to inclusive breakup cross sections \cite{Shri06, Pand16, Chatto18b, Jha20}.  It has been shown experimentally that the breakup through the indirect path, namely the  process consisting of transfer followed by breakup may provide a dominant contribution \cite{Shri06, Pand16}. Apart from the breakup process that is related to the low $\alpha$-binding energies,  the importance of neutron transfer has been emphasized for the large production of inclusive $\alpha$ cross section \cite{Prad13, Parkar21, Jha20}. While many studies have focused on the contribution of breakup process, the role of neutron transfer has not been investigated well enough. There are not many measurements for the  neutron transfer cross section with WBPs and data are scarce. \cite{Shri06, VVP18, Prad13, Hu16, Fang16, Fisi17, Zhang18, Pandit17, Pals14, Ara13}. 

Due to availability of Radioactive Ion Beam (RIB), the features observed in stable WBPs can be explored also for the nuclei away from the line of stability \cite{Keel09, Keel07, Kola16, Cant15}. In these nuclei, besides the low binding energy and cluster structure, there exists a long tail in the density distribution corresponds to anomalously large size. This may be interpreted in terms of halo and borromean structures (nuclei comprising three bound components in which any subsystem of two components is unbound). In many of these nuclei, one or many neutron transfer are found to give quite dominant contribution to the reaction cross section. For example, in the study of $^{6,8}$He+$^{65}$Cu, $^{197}$Au systems, 2n transfer channel was shown to be dominant \cite{Chat08, Lema10, Lema11}. 

On the theoretical front, coupled channel calculations have been successfully utilised for explaining the most of the experimental observables and to elucidate on the underlying reaction mechanism. The Continuum Discretized Coupled Channels (CDCC) method has been used to study the breakup process by including the couplings to the continuum  states above the breakup threshold of projectile nucleus. By this method, both the one-step process of continuum excitation called as prompt breakup and the two-step process of excitation of long-lived resonances in the continuum followed by decay, namely the  delayed breakup are taken into account. The transfer processes are described through the Coupled Reaction Channel (CRC) calculations that involves the multi-state couplings in the projectile and target like nucleus before and after the transfer of nucleon(s). 

While the systematics of the fusion and elastic-scattering have been studied well, the systematic studies of the available data of neutron transfer have not been performed.  In the present work, we study the one-neutron stripping and pickup cross sections in $^{7}$Li+$^{124}$Sn reaction and investigate the relative importance of the breakup and transfer processes through the CDCC and CRC calculations. The systematics of neutron transfer cross sections for different targets is also studied. The paper is organized as follows: Calculation details are given in section \ref{sec:calc}. The results are discussed in section \ref{sec:results} and summary is given in section \ref{sec:Sum}.

\section{\label{sec:calc} Calculation details}
To understand the mechanism of transfer and breakup reactions, coupled channel calculations have been performed. We have performed three kind of calculations (i) CRC using phenomenological global optical model potentials (iii) CRC using normalised microscopic S$\tilde{a}$o Paulo potentials, (iii) CDCC and (iv) CDCC+CRC. All these calculations have been performed using the code FRESCO (version FRES 2.9) \cite{Thom88}. Next, we discuss about the calculation methods in detail.

\subsection{CRC Calculations}
In these type of calculations, optical model potentials for entrance and exit channels are required. In the following subsections, we discuss calculations using phenomenological global optical model potentials and microscopic double folding model potentials. Apart from this, binding potentials of the fragment and core for the projectile and target partitions are required. The potentials binding the transferred particles were of Woods-Saxon volume form, with radius 1.25A$^{1/3}$ fm and diffuseness 0.65 fm, with `A' being the mass of the core nucleus. The depths were automatically adjusted to obtain the required binding energies of the particle-core composite system. The single particle states along with Spectroscopic factors (C$^2$S) considered in the calculations are given in Table\ \ref{SA}. For the $^7$Li $\rightarrow$ $^6$Li+$n$ transfer, both the 1p$_{3/2}$ and 1p$_{1/2}$ components of the neutron bound to $^6$Li were included with spectroscopic factors of C$^2$S = 0.43 and 0.29 respectively, taken from Cohen and Kurath \cite{Cohe67}. Similarly for $^7$Li+$n$ $\rightarrow$ $^8$Li transfer, both the 1p$_{3/2}$ and 1p$_{1/2}$ components of the neutron bound to $^7$Li were included with spectroscopic factors of C$^2$S = 0.98 and 0.056 respectively, taken from Cohen and Kurath \cite{Cohe67}. The finite range form factors in the post form for stripping and prior form for pickup were used. Calculations were carried out including the full complex remnant term. 
   
\subsubsection{CRC Calculations using global phenomenological optical model potentials}
Recently, the global phenomenological optical model potentials for $^6$Li, $^7$Li and $^8$Li have been proposed \cite{6Li_Xu18, 7Li_Xu18, 8Li_Su17} which have been used for entrance $^7$Li+$^{124}$Sn and exit $^6$Li+$^{125}$Sn (for stripping) and $^8$Li+$^{123}$Sn (for pickup) channels. We refer the results of these calculations as CRC1. The potential parameters are listed in Table\ \ref{tab:potential}. The old set of global phenomenological optical model potentials for $^{6,7}$Li of Cook $\textit{et al.}$ \cite{Cook82} have also been tried for comparison and the results have been found to be the similar.  

\subsubsection{CRC Calculations using S$\tilde{a}$o Paulo potentials}
The calculations have also been performed using microscopic double-folding S$\tilde{a}$o Paulo potentials \cite{Chamon97, Chamon02} for real and imaginary parts of the optical potential. At near barrier energies, this potential is equivalent to the usual double folding potential with the advantage that it has a comprehensive systematic for the matter densities. For this reason, this can be considered as a parameter-free potential. Since the breakup channel was not considered explicitly in CRC calculations, the strength coefficients for real and imaginary potentials were kept as N$_R$ = N$_I$ = 0.6. Similar method was adopted in Refs.\ \cite{Pere09, Sousa10, Saku83, Hu16, Fang16} to account for the loss of flux to dissipative and breakup channels \cite{Pere09, Sousa10} and repulsive nature of the real part of the breakup polarization potential \cite{Mac09, Kumawat08, Parkar11, Santra11, Parkar13}. In the outgoing partition, the S$\tilde{a}$o Paulo potential was used for both the real and the imaginary parts with strength coefficients N$_R$ = 1.0 and N$_I$ = 0.78. This procedure has been shown to be suitable for describing the elastic scattering cross section for many systems in a wide energy interval \cite{Gasq06}. We refer these calculations as CRC2.

\subsection{CDCC and CDCC+CRC calculations}
To investigate the effect of projectile breakup and neutron transfer on elastic scattering simultaneously, the CDCC and CDCC+CRC calculations have been carried out. Both the inelastic (bound and unbound) excitations of the projectile and neutron transfer channels have been coupled.

The coupling scheme used in CDCC is similar to that described in earlier works \cite{Parkar08, Jha09}. The calculations assumed a two-body $\alpha-t$ cluster structure for the $^7$Li nucleus. The ground state and inelastic excitation of $^7$Li were considered as pure L = 1 cluster states, where L is the relative angular momentum of clusters. The continuum above the $^7$Li$\rightarrow\alpha$+t breakup threshold (2.47 MeV) was discretized into bins of constant momentum width $k = 0.20$ $fm^{-1}$, where $\hbar k$ is the momentum of $\alpha + t$ relative motion. The binding potentials for all the bound and continuum cluster states were the well-known potentials from Ref.\ \cite{Buck88}. The cluster wave functions for each bin in the continuum were averaged over the bin width and each of these bins was then treated as an excited state of $^7$Li with an excitation energy equal to the mean of the bin energy range. The continuum momentum bins were truncated at the upper limits of $k_{max} = 0.8$ $fm^{-1}$ for the calculations. The continuum states with relative orbital angular momentum L = 0, 1, 2, and 3 were included. In addition the full continuum continuum (CC) couplings were taken into account in the final calculations. The real part of required fragment-target potentials V$_{\alpha-T}$ and V$_{t-T}$ in cluster folding model were taken from S$\tilde{a}$o Paulo potential \cite{Chamon02}, while short range imaginary potential with values W$_0$ = 25 MeV, r$_w$ = 1.00 fm, a$_w$ = 0.40 fm was used. In addition to CDCC calculations for breakup, the CRC calculations of type CRC1 as explained above were simultaneously performed.

\begin{table*}
\centering
\caption{Optical model potential parameters used in CRC calculations. The radius parameter in the potentials are derived from R$_i$=r$_i$.A$^{1/3}$, where i = R, V, S, C and A is the target mass number.}\label{tab:potential}
\begin{tabular}{cccccccccccc}
\hline
System&V$_R$ (MeV)& r$_R$ (fm)& a$_R$ (fm)& W$_V$ (MeV)& r$_V$ (fm)& a$_V$ (fm)& W$_S$ (MeV)& r$_S$ (fm)& a$_S$ (fm)& r$_C$ (fm)& Ref. \\
\hline
$^7$Li+$^{124}$Sn & 179.9 & 1.24 & 0.85 & 22.22 & 1.59 & 0.60 & 36.01 & 1.18 &  0.87 & 1.80 & \cite{7Li_Xu18} \\
$^6$Li+$^{125}$Sn& 259.2 & 1.12 & 0.81 & 0.49 &  1.54 & 0.73 & 25.29 & 1.31 &  0.94 & 1.67 & \cite{6Li_Xu18} \\
$^8$Li+$^{123}$Sn & 109.5 & 1.33 & 0.81 & 29.03 & 1.53 & 0.88 & & & & 1.57 & \cite{8Li_Su17} \\
\hline
\end{tabular}
\end{table*}

\begin{table}
\begin{center}
\caption{\label{SA} Energy levels of residual nuclei and spectroscopic factors (C$^2$S) used for neutron transfer channels : $^{124}$Sn $\rightarrow$ $^{125}$Sn \cite{Bingh73} and $^{124}$Sn $\rightarrow$ $^{123}$Sn \cite{Flem82}.}\ \\
\begin{tabular}{|ccc|ccc|}
\hline \multicolumn{3} {|c|} {$^{125}$Sn}&
\multicolumn{3} {|c|} {$^{123}$Sn} \\ \hline
E & J$^{\pi}$ & C$^2$S & E& J$^{\pi}$ & C$^2$S \\
(MeV)& & & (MeV)& & \\ \hline
0.000 & 11/2$^-$ & 0.42 &0.000 &$11/2^-$ &4.49 \\
0.026 & 3/2$^+$ & 0.44 &0.025 &$3/2^+$ &4.49 \\
0.232 & 1/2$^+$ & 0.33 &0.139 &$1/2^+$ & 1.90\\
0.930 & 7/2$^-$ & 0.015 &0.920 &$3/2^+$ & 1.00\\
1.277 & 5/2$^+$ & 0.07 &1.028 &$7/2^+$ & 2.79\\
1.377 & 7/2$^+$ & 0.038 &1.484 &$5/2^+$ & 2.79\\
1.555 & 5/2$^+$ & 0.040 &1.155 &$7/2^+$ & 3.20\\
2.264 & 5/2$^+$ & 0.019 &1.194 &$5/2^+$ & 1.00\\
2.600 & 7/2$^-$ & 0.010 &1.784 &$5/2^+$ & 1.00\\
2.767 & 7/2$^-$ & 0.54 &1.902 &$5/2^+$ & 1.00\\
2.890 & 7/2$^-$ & 0.032 &2.026 &$5/2^+$ & 1.00\\
3.016 & 7/2$^-$ & 0.040 &2.365 &$7/2^+$ & 1.00\\
3.085 & 7/2$^-$ & 0.040 &2.446 &$1/2^+$ & 1.00\\
&&&2.850&$5/2^+$&1.00\\
\hline
\end{tabular}
\end{center}
\end{table}

\section{Results and Discussion}\label{sec:results}
\subsection{Elastic Scattering}
\begin{figure}
\includegraphics[width=85mm,trim=2.6cm 10.7cm 8.5cm 9.8cm, clip=true]{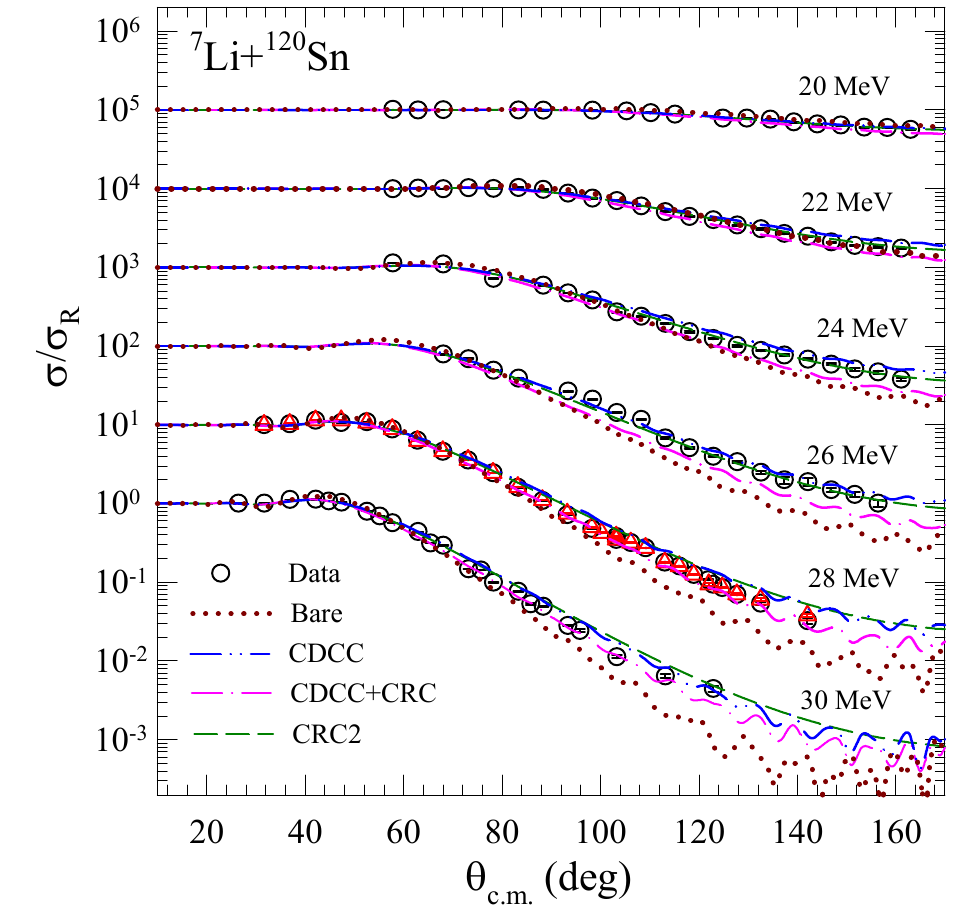}
\caption{\label{elastic} Elastic scattering data for $^7$Li+$^{120}$Sn system \cite{Kundu17, Zaga17} is compared with the calculations. Red triangle data is for $^7$Li+$^{124}$Sn system \cite{Kundu19} at 28 MeV (see text for details).}
\end{figure}
\begin{figure}
\includegraphics[width=85mm,trim=0.6cm 19.8cm 3.8cm 2.6cm, clip=true]{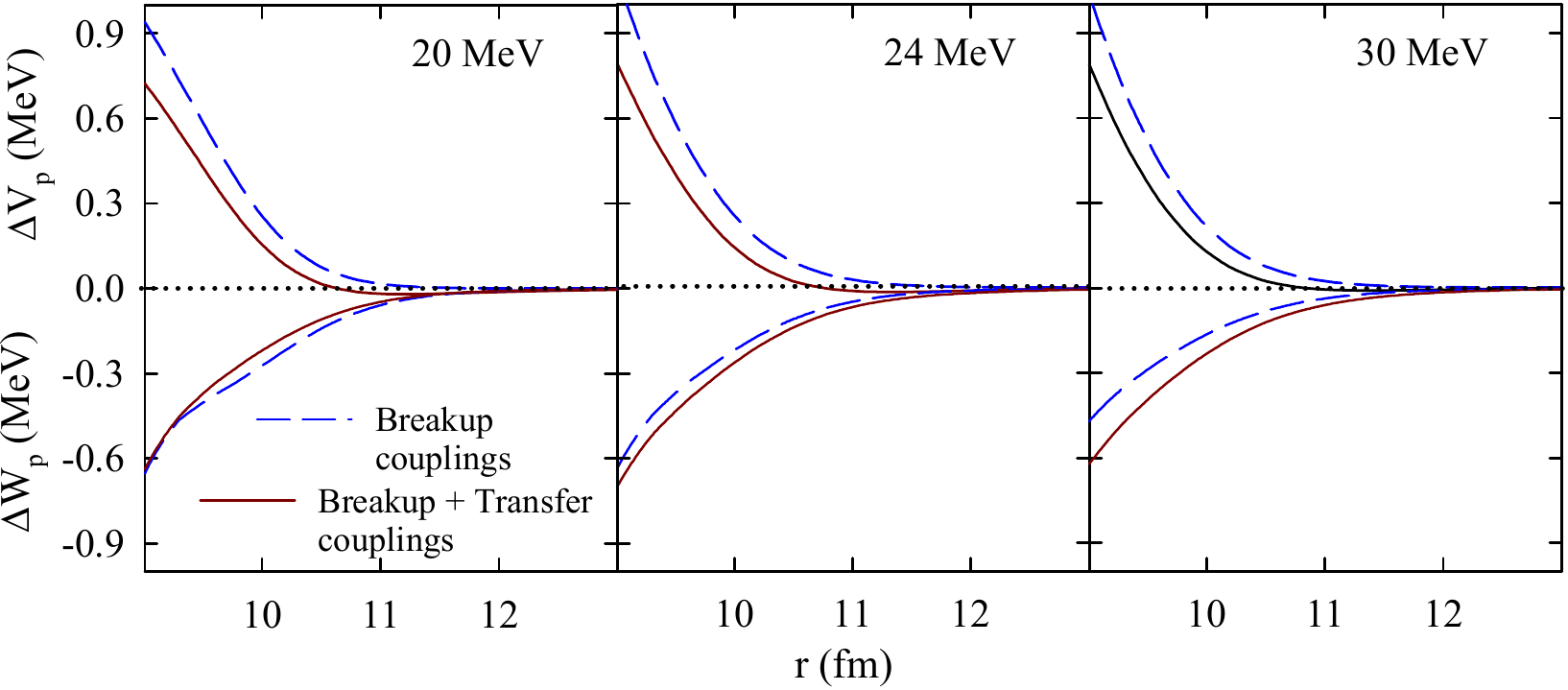}
\caption{\label{PolPot} Real and imaginary parts of dynamic polarisation potentials due to breakup (dashed lines) and breakup+transfer (solid lines) couplings in $^7$Li+$^{124}$Sn system at three bombarding energies 20, 24 and 30 MeV.}
\end{figure}

The elastic scattering data available for $^7$Li+$^{124}$Sn system at 28 MeV \cite{Kundu19} and for $^7$Li+$^{120}$Sn at 20, 22, 24, 26, 28, and 30 MeV \cite{Kundu17, Zaga17} was utilised for testing our entrance channel potentials and also to see the effect of breakup and neutron transfer couplings on the elastic scattering angular distributions. It is to be noted that the elastic scattering data with $^7$Li projectile on $^{120}$Sn and $^{124}$Sn targets at 28 MeV was similar as shown in Fig.\ \ref{elastic}. The calculations of CRC2 type along with CDCC and CDCC+CRC are shown along with the data in Fig.\ \ref{elastic}. Dotted lines are the bare calculations without including any continuum couplings. The coupling effects are evident at above barrier energies. Around barrier (20 and 22 MeV), the coupling effects are negligible. 

To understand the observation from the coupling effects in the calculations for the elastic scattering angular distribution in a better way, we have investigated the behaviour of the dynamic polarisation potential (DPP) generated due to these couplings. The calculated DPPs due to breakup (CDCC calculation) and breakup+transfer (CDCC+CRC calculations) couplings in the vicinity of the strong absorption radii are shown for three (20, 24 and 30 MeV) energies in Fig.\ \ref{PolPot}. It is evident from Fig.\ \ref{PolPot} that the breakup couplings give rise to repulsive real and attractive imaginary DPPs. After inclusion of transfer couplings, the real part of DPP is slightly reduced. Similar behaviour was also observed in the reaction with $^9$Be projectile \cite{Parkar13}.

\subsection{1n stripping and 1n pickup}
\begin{figure}[htbp]
\includegraphics[width=85mm,trim=1.4cm 20.3cm 10.5cm 2.0cm, clip=true]{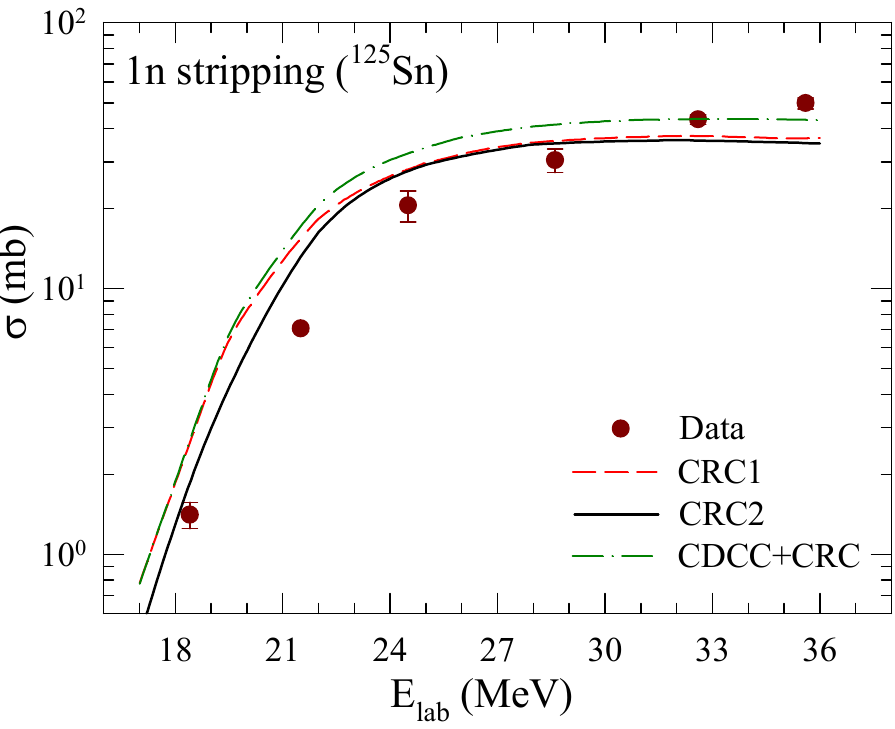}
\caption{\label{1nstripping} Measured one neutron stripping cross sections in $^7$Li+$^{124}$Sn system are compared with the calculations.}
\end{figure}

The 1n stripping data leading to $^{125}$Sn channel was measured in Ref.\ \cite{VVP18} by offline $\gamma$ counting. We have compared this data with the present set of calculations (CRC1, CRC2 and CDCC+CRC) and results are shown in Fig.\ \ref{1nstripping}. A good agreement between the data and calculations imply that the states in residual nucleus $^{125}$Sn upto 3 MeV (given in Table\ \ref{SA}) that were included in the calculations are sufficient to explain the measured data reasonably well. Contribution from the states higher than 3 MeV or other indirect paths do not have a significant role in the description of experimental data.

\begin{figure}[htbp]
\includegraphics[width=85mm,trim=1.4cm 20.5cm 10.0cm 1.4cm, clip=true]{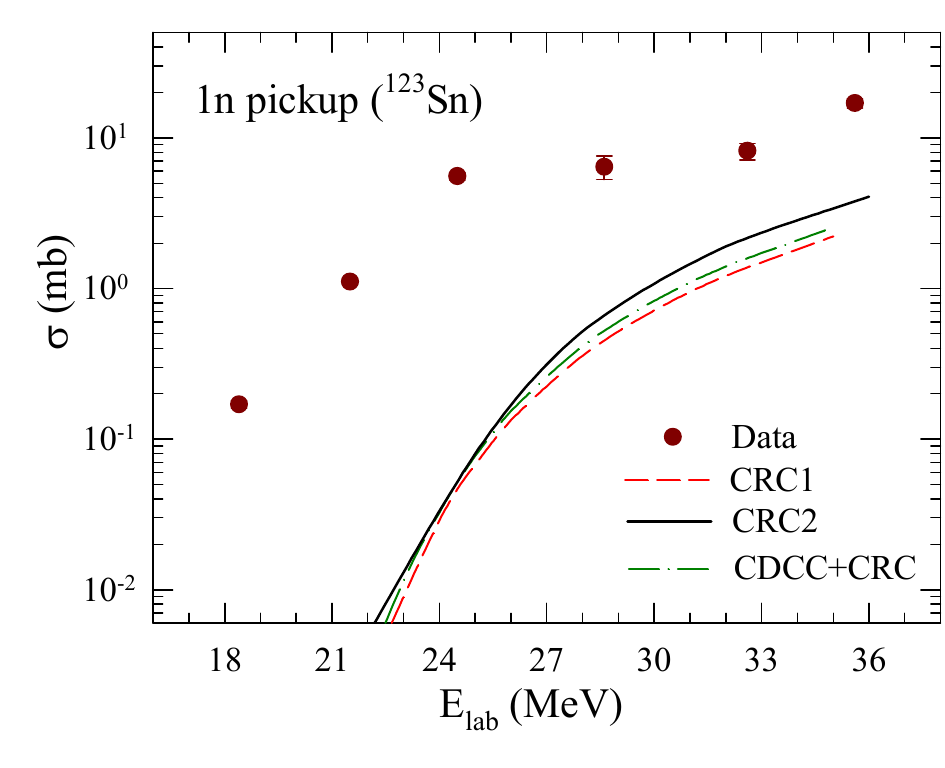}
\caption{\label{1npickup} Measured one neutron pickup cross sections in $^7$Li+$^{124}$Sn system are compared with the calculations.}
\end{figure}
Similar to 1n stripping channel the data for 1n pickup leading to $^{123}$Sn residual nucleus was also measured in Ref.\ \cite{VVP18}. We have compared this data also with the present set of calculations (CRC1, CRC2 and CDCC+CRC). The data and the calculation results are shown in Fig.\ \ref{1npickup}. The states in residual nucleus $^{123}$Sn up to 3 MeV (given in Table\ \ref{SA}) were coupled. The spectroscopic information about only few states having large spectroscopic factors are available in the literature \cite{Flem82}.  Since the spectroscopic information was only partially available for the  $^{123}$Sn residual nucleus, the calculations show a large under prediction as compared to the data as seen in Fig.\ \ref{1npickup}. In addition, there might be contributions from other indirect paths to this channel.

\subsection{Systematics of neutron transfer cross sections with $^7$Li projectile}
\begin{figure*}
\includegraphics[width=160mm,trim=1.1cm 20.9cm 3.4cm 2.4cm, clip=true]{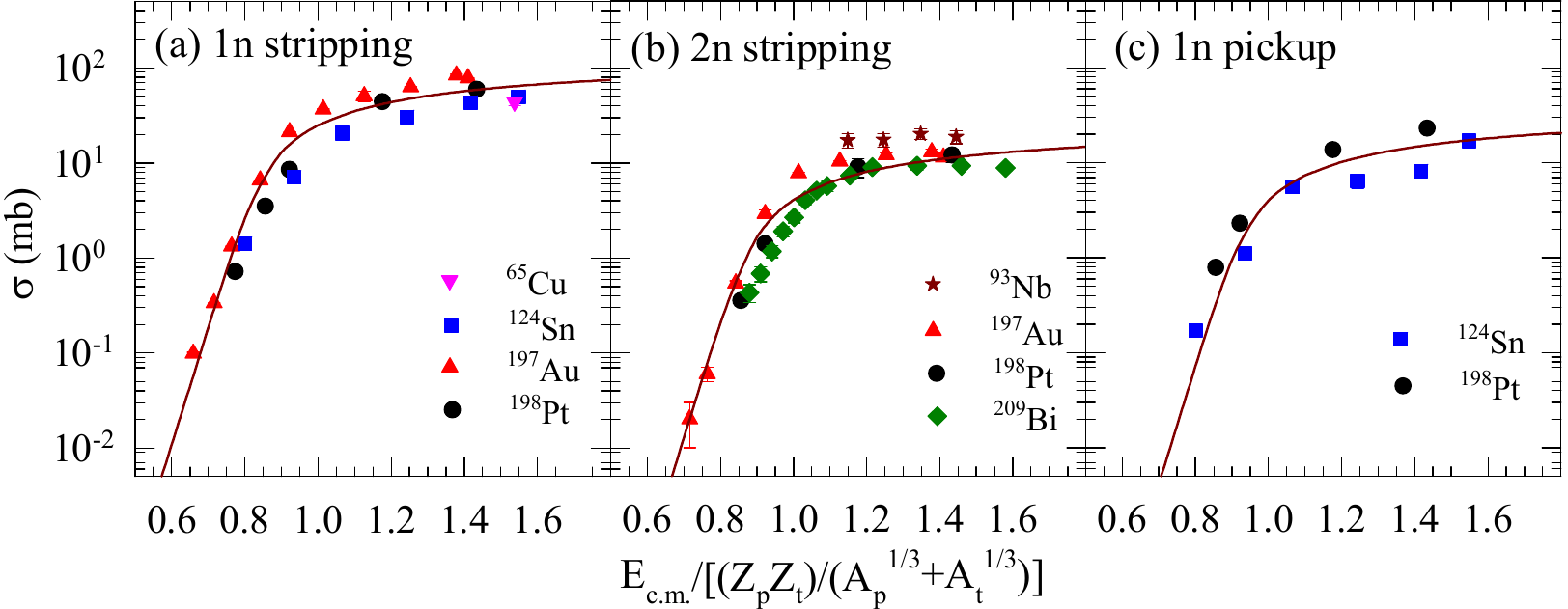}
\caption{\label{7Li_transfer}Systematic behaviour of (a) one neutron stripping, (b) two neutron stripping, and (c) one neutron pickup cross sections as a function of reduced energy with $^7$Li projectile on various targets. Lines are fit to the data (see text for details). }
\end{figure*}
\begin{figure}
\includegraphics[width=85mm,trim=1.5cm 17.2cm 9.5cm 2.7cm, clip=true]{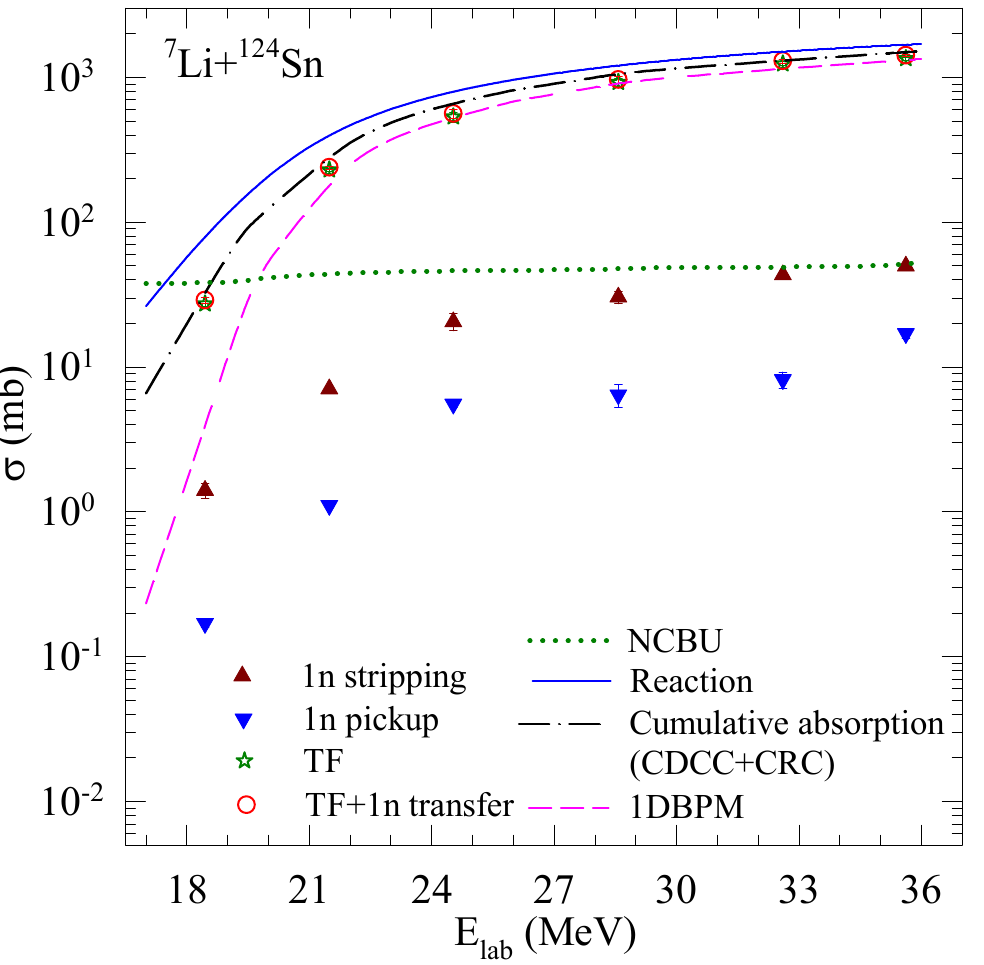}
\caption{\label{reactionmech} Measured cumulative TF and neutron transfer cross sections are compared with the reaction cross sections. NCBU, cumulative absorption and BPM model calculations are also shown (see text for details).}
\end{figure}

The data available for 1n stripping, 2n stripping and 1n pickup cross sections with $^7$Li projectile on $^{65}$Cu \cite{Shri06}, $^{93}$Nb \cite{Pandit17}, $^{124}$Sn \cite{VVP18}, $^{197}$Au \cite{Pals14}, $^{198}$Pt \cite{Ara13} and $^{209}$Bi \cite{Dasgupta04} targets is plotted in Fig.\ \ref{7Li_transfer}(a-c). The variable on X-axis is chosen so as to remove any geometrical factors due to target size. As can be seen from the figure, universal behaviour in the cross sections in all the three plots are observed. The neutron transfer systematics are quite interesting and it is the first time that these systematic have  been presented with the $^7$Li projectile. With the $^9$Be projectile, similar universal behaviour in neutron stripping cross sections was observed \cite{Fang16}. In addition, similar universal behaviour was earlier also shown for the inclusive $\alpha$ \cite{Santra12, Pandit17, Jha20}, triton capture \cite{Pandit17}, fusion \cite{Jha20} and reaction \cite{Kola16} cross sections. In order to explain the appearance of universal barrier in these plots Fig.\ \ref{7Li_transfer}(a-c), we have used the Wong formula \cite{Wong73} that is based on the barrier penetration and we have multiplied it by transfer probability (\(exp(-c.S_{n/2n}))\) as given below.  

\begin{eqnarray}
\sigma=\frac{\hbar\omega}{2E_{c.m.}}R_b^2\log\left[1+exp\left(\frac{2\pi}{\hbar\omega}(E_{c.m.}-V_b-a)\right)\right] \\ \nonumber 
 exp(-cS_{n/2n})
\end{eqnarray}
where \textit{a} and \textit{c} are the parameters that were varied to fit the data. S$_n$ and S$_{2n}$ are the separation energies for 1n stripping / pickup and for 2n stripping, respectively . Parameter `$a$' represents the shift in the barrier for the specific reaction channel  while parameter `$c$' provides the overall normalization with respect to fusion cross sections to describe the transfer cross section in magnitude. The values of V$_b$, R$_b$ and $\hbar\omega$ for $^7$Li+$^{124}$Sn were used from Ref.\ \cite{VVP18}. The resulting fits are shown as the solid lines in Fig.\ \ref{7Li_transfer}(a-c). The values of \textit{a} and \textit{c} that have been obtained are given in Table\ \ref{acvalues}. 
\begin{table}
\centering
\caption{$a$ and $c$ values obtained from the fitting of universal plots of Fig.\ \ref{7Li_transfer}.}\label{acvalues}
\begin{tabular}{cccc}
\hline
Process & S$_{n/2n}$ (MeV) & a (MeV)& c (MeV$^{-1}$) \\
\hline
(a) 1 n stripping & 7.25 & -3.58 & 0.44  \\
(b) 2 n stripping & 12.92 & -2.86 & 0.37  \\
(c) 1 n pickup & 2.03 & -1.86 & 2.16  \\
\hline
\end{tabular}
\end{table}
These values of parameters $a$ and $c$ explain the early onset and larger magnitude of the 1n transfer as compared to other two channels observed in the data.

\subsection{Reaction mechanism in $^7$Li+$^{124}$Sn system}
To understand the complete reaction mechanism in $^7$Li+$^{124}$Sn system, the measured TF and neutron transfer cross sections \cite{VVP18} have been compared with deduced reaction cross sections from the present calculations in Fig.\ \ref{reactionmech}. We have also plotted the fusion cross sections calculated by the barrier penetration model (BPM) method, which reproduces the experimental TF at above barrier energies but under predict it at sub-barrier energies. The cumulative absorption cross sections from CDCC+CRC calculations are found to agree with the cumulative TF and neutron transfer cross sections. Non-capture breakup (NCBU) calculated from CDCC is also shown, which contributes higher than 1n transfer at below barrier energies. From a comparison between the reaction cross sections and the sum of TF and neutron transfer cross sections, it can be concluded that other reaction channels, such as, proton transfer channels, non-capture breakup and target inelastic states may also contribute to the reaction cross section for which there are no measurements. 

\section{\label{sec:Sum} Summary}
We have investigated the important underlying reaction mechanisms, namely breakup and neutron transfer for the $^7$Li+$^{124}$Sn system around the Coulomb barrier energies. These processes are found to affect the elastic scattering and the fusion cross sections. We have performed the CRC calculations using the global optical model potential parameters as well as using the S$\tilde{a}$o Paulo potential for transfer studies for the $^7$Li+$^{124}$Sn system. We have further performed CDCC+CRC calculations to take into account the combined effects of breakup and transfer channels. The simultaneous explanation of elastic scattering, 1n stripping and total fusion was obtained from these calculations. One of the important findings of this work is the explanation of the observed universal behaviour of stripping and pickup cross sections with $^7$Li projectile on several targets. It shows the early onset of 1n transfer compared to fusion reactions and other transfer reaction channels. It is observed that while 1n stripping contributes significantly, other reaction channels such as, proton transfer, non-capture breakup and target inelastic states might be necessary for a complete description of the reaction cross section for the $^7$Li+$^{124}$Sn system. 

\begin{acknowledgments}
The authors V.V.P. and S.K. acknowledge the financial support from Young Scientist Research grant and Senior Scientist programme, respectively, from the Indian National Science Academy (INSA), Government of India, in carrying out these investigations.
\end{acknowledgments}

\end{document}